# Role of HIV RNA structure in recombination and speciation: romping in purine A, keeps HTLV away

**Donald R. Forsdyke**

*Department of Biomedical and Molecular Sciences, Queen's University, Kingston, Ontario, Canada K7L3N6*

Corresponding author at: Department of Biomedical and Molecular Sciences, Queen's University, Kingston, Ontario, Canada K7L3N6. Tel.: 613-533-2980 Fax.: 613-533-6796

*E-mail address:* forsdyke@queensu.ca

ABSTRACT

Extreme enrichment of the human immunodeficiency virus (HIV-1) RNA genome for the purine A parallels the mild purine-loading of the RNAs of most organisms. This should militate against loop-loop "kissing" interactions between the structured viral genome and structured host RNAs, which can generate segments of double-stranded RNA sufficient to trigger intracellular alarms. However, human T cell leukaemia virus (HTLV-1), with the potential to invade the same host cell, shows extreme enrichment for the pyrimidine C. Assuming the low GC% HIV and the high GC% HTLV-1 to share a common ancestor, it was postulated that differences in GC% arose to prevent homologous recombination between these emerging lentiviral species. Sympatrically isolated by this intracellular reproductive barrier, prototypic HIV-1 seized the AU-rich (low GC%) high ground (thus committing to purine A rather than purine G). Prototypic HTLV-1 forwent this advantage and evolved an independent evolutionary strategy. Evidence supporting this hypothesis since its elaboration in the 1990s is growing. The conflict between the needs to encode accurately both a protein, and nucleic acid structure, is often resolved in favour of the nucleic acid because, apart from regulatory roles, structure is critical for recombination. However, above a sequence difference threshold, structure (and hence recombination) is impaired. New species can then arise.

## 1.Recombination, rejuvenation and conservation of RNA structure

At its most fundamental level, sex is recombination (Forsdyke, 2007; Michod et al., 2008). The 'romp in the hay' that can precede the meeting of paternal and maternal human gametes is an elaborate, albeit necessary, preliminary to the final meiotic meeting of parental genomes in the gonads of their offspring, where recombination occurs. The early microscopists who witnessed this final meeting described it as a "conjugation of the chromosomes" that was necessary for a rejuvenating "interchange of substances" (Montgomery, 1901):

> "The conjugation of the chromosomes in the synapsis stage may be considered the final step in the process of conjugation of the [parental] germ cells. It is a process that effects the

> rejuvenation of the chromosomes; such rejuvenation could not be produced unless chromosomes of different parentage joined together, and there would be no apparent reason for chromosomes of like parentage to unite."

In contrast to humans, retroviral sex is quite elementary (Temin, 1991). Yet, like humans, retroviruses are diploid. This diploidy in HIV-1 can be heterozygous due to the viral strategy of mutation to near oblivion, so countering host defences. The degree of heterozygosity in two HIV genomes that are copackaged within an infectious retrovirus particle, if below the sequence difference threshold above which recombination is inhibited (see Section 7), will allow viral rescue by recombination. Thus, within the next host cell, the two partially crippled genomes repeatedly recombine to generate a rejuvenated form that will successfully colonialize the vulnerable population of host cells (usually T4 lymphocytes; Smyth et al., 2012).

Apart from this, HIV genomes literally 'romp' in the purine A. The highly stable A-richness of certain lentivirus genomes (Kypr and Mrazek, 1987), was recently reviewed by Kuyl and Berkhout (2012). Noting the growing evidence that "RNA genome structure, and not the encoded proteins, is the most decisive factor that triggers HIV conservation" (Forsdyke, 1995; Watts et al., 2009; Simon-Loriere et al., 2010; Snoeck et al., 2011; Sanjuán and Borderia, 2011), they suggested that A-richness would facilitate the reverse transcription stage of the viral life cycle by ensuring that such RNA structures are of low stability (since A-U bonds are relatively weak). Indeed, as might be expected, lowering the A-content of codons without changing the encoded amino acids, can impede cDNA synthesis (Keating et al., 2009). As for RNA structure itself, it was held that this would confer a selective advantage by providing "packaging signals." Thus, the viral nucleocapsid packaging protein could be considered a "sensor of nucleotide composition." But, paradoxically, "genes … from multiple sources (e.g. bacterial, viral, and human) are tolerated by … HIV-based vectors, suggesting that nucleotide composition of the insert … is not critical for packaging" (Kuyl and Berkhout, 2012).

There have long been other views on these issues. Viral structure has been related to viral recombination and speciation (Forsdyke, 1995, 1996; Forsdyke, 2001). Viral A-richness has been related to the phenomenon of purine-loading (Forsdyke and Mortimer, 2000). These are the subjects of this paper. In keeping with current conventions (Forsdyke, 2011a), "high A%" implies a high frequency of the base A. "High AT%" implies a high frequency of the combination of bases A and T, without specification of which of the two bases makes the major contribution. The same considerations apply to high values for GC% and AG% ("purine loading").

## 2. Purine-loading

The A-bias in HIV-1 is a characteristic of coding regions, but not of the non-coding long terminal repeats (LTRs). This parallels the selective "purine-loading" (AG% enrichment) of the encoding (mRNA synonymous) strands of the exons of genes of many biological species, including man. Since purines do not pair with purines, this would militate against self RNA-self RNA interactions, and a purine-loaded virus would be less likely to form base-paired RNA duplexes with host RNAs. Preventing dsRNA formation should be advantageous to the virus, since segments of double-stranded RNA (dsRNA) can provide alarm signals to the host, so alerting immunological defences (Cristillo et al., 2001). Base pairing being largely entropy-driven, purine-loading should be high in thermophiles, as is indeed found (Lao and Forsdyke, 2000; Lambros et al. 2003).

Purine-loading also provides a rationale for interruption of proteins by low complexity segments corresponding to purine-rich codons (Xue and Forsdyke, 2003; Tian et al., 2011). This implies function at the nucleic acid level. Thus, Muralidharan et al. (2011), from studies of a *P. falciparum* protein with an asparagine repeat sequence (corresponding to purine-rich codons GAY), concluded that "the asparagine repeat is dispensible for protein expression, stability and function." A similar observation has been made for a purine-rich repeat within the mRNA encoding the EBNA-1 protein, which normally maintains the Epstein-Barr virus (EBV) in a deeply latent state with its many pyrimidine-rich mRNAs unexpressed. In this way EBV should evade host immune defences (Cristillo et al., 2001; Starck et al., 2008; Tellam et al., 2012).

Irrespective of whether avoidance of dsRNA is the correct adaptive explanation, purine loading is a fact. A-loading by HIV-1 could be an extreme expression of this (Cristillo et al., 2001). Purine-loading can influence amino acid composition and, since the extra purines are generally located at third codon positions (the main site for synonymous mutations), it is likely that purine-loading has driven amino acid composition, rather than the converse (Mortimer and Forsdyke, 2003). The purines would be expected to occupy, and are indeed found in, the loop regions of the highly structured HIV genome (Forsdyke and Bell, 2004; Watts et al., 2009). Indeed, Hemert et al. (2013) confirm for HIV-1 RNA that the percentage of A nucleotides is particularly low in double-stranded structures (21%) compared to the single stranded parts (79%), and suggest that “the virus may have adopted a particular genome architecture to adapt to cellular defense mechanisms.” Furthermore, Zanini and Neher (2013) note that most synonymous (non-amino acid changing) mutations disrupt base pairing in RNA stems, so “hinting at a direct fitness effect of these stem-loop structures.”

Values for AG% and GC% are reciprocally related, and purine-loading is achieved mainly by replacing C with A, leaving T and G relatively constant (so avoiding runs of Ts and Gs that can be detrimental). Thus, there can be conflict between ‘AG pressure’ and ‘GC-pressure’ (Mortimer and Forsdyke, 2003; Forsdyke, 2011a). To fully appreciate why HIV-1 is A-loaded, not G-loaded, we must understand the role of GC%.

**3. Two species in one cell differ greatly in GC%**

Consider three examples of pairs of viral species, where members of each pair might occupy the same host cell. (1) An early report on species differences in GC% (Wyatt, 1952) noted that, while there is no clear relationship between the GC% values of virus and host, two insect viruses (polyhedral and capsule) with a common larval host (spruce budworm) differ dramatically in their GC% values (38% versus 51%). In this example it is uncertain whether these viruses could in fact have shared a common cell within their larval host.

(2) However, Schachtel et al. (1991) noted for humans, that two neurotropic alphaherpesviruses also differ dramatically in GC% (46% versus 68%). The differences are dispersed, being present in both non-coding and coding regions and, in the latter case, mainly affect the third bases of codons (consistent with the close similarity in amino acid sequences of the viral proteins). But, as in the case of purine-loading (Section 2), the differences are sometimes sufficient to change amino acids (e.g. lysine to arginine). It was proposed that the differences are adaptive in that there would be less competition for host resources:

> “More specifically, it is proposed that HSV1 and VSV avoid competition for host resources such as nucleic acid precursors and aminoacyl-tRNAs through divergent base composition and codon usage. This may enable these two closely related viruses to coexist in the same host species and even to multiply simultaneously in the same cells."

(3) A similar explanation was advanced by Bronson and Anderson (1994) who noted that HIV-1 is extremely AU-rich, and HTLV is extremely GC-rich. They supposed that the intracellular environment would consist of a variety of metabolic "ecological niches." Two viruses that coinfected the same host cell would avoid competition with each other by making different metabolic demands on NTP pools. There is, however, another interpretation.

## 4. Coinfectants either blend or speciate

The wide GC% differences between coinfecting viruses can be seen as part of the speciation process (Forsdyke, 1995, 1996, 2001). Two species of virus derived from a common ancestral species, which coexist synchronously within the same host cell (sympatry), have the opportunity to recombine ('blending inheritance'). But their reproductive isolation would then be lost and they would mutually destroy each other as independent species. To the extent that retaining their differentiation as independent species is advantageous, there would have been a strong *selective pressure* on their nucleic acids to evolve to prevent recombination. If differences in GC% could achieve this (Section 7), then they would have been mutually driven to extreme GC% poles. If HIV-1 were the first to A-load, then it would have been driven to the low GC% pole. HTLV would have been driven to the high GC% pole. Since, to arrive at high GC% values, organisms preferentially replace A with C (Section 2), HTLV would have foregone the advantages of purine-loading and adopted an alternative evolutionary strategy (Cristillo et al., 2001). With its 'romp' in the A, HIV-1 recombinationally drove HTLV-1 away!

Like the GC-rich EBV (Section 2), HTLV-1 is deeply latent and many carriers are symptom-free. Also like EBV, the duplex HTLV-1 provirus avoids transcribing its many pyrimidine-rich RNAs (controlled by the 5' LTR region), by expressing a protein (HBZ) with a role analogous to that of the EBNA-1 protein. HBZ mRNA is transcribed from the antisense strand as a purine-rich, RNA species (controlled by the 3' LTR; Cook et al., 2013).

## 5. FORS-D Analysis and SHAPE illuminate RNA structure

New technologies have greatly assisted the understanding of HIV RNA structure. Le and al. (1988, 1989) used novel energy-minimization folding algorithms to show that RNA regions involved in intra-molecular base pairing tend to be more evolutionary stable. This stability of nucleic acid folding into stem-loop conformations is expected to be influenced by base composition, being low in low GC% organisms such as HIV-1 (since there is less opportunity for pairing between Gs and Cs). However, GC% values tend to characterize whole genomes or large genome sectors. Base order is a character that critically affects local structure. A method of dissecting out the base order-dependent component of the folding energy – "folding of randomized sequence difference" (FORS-D) analysis – revealed many evolutionarily conserved structural elements in the HIV RNA genome; these were separated by less structured variable regions likely to be under positive Darwinian selection (Forsdyke, 1995; Snoeck et al., 2011; Zanini and Naher, 2013).

These observations were confirmed and extended by "selective 2'-hydroxyl acylation analysis by primer extension" (SHAPE; Watts et al., 2009). While certain conserved structures corresponded to established regulatory elements, others corresponded to the peptide linkers between the protein domains of polyproteins. The results were interpreted in terms of "another level of genetic code" so that "higher ordered RNA structure directly encodes protein structure." Recombination was not mentioned. Others agreed, stating that "this novel component of the genetic code represents the strongest determinant of conservation" (Snoeck et al., 2011). Again, there was no suggestion of a relationship to recombination.

## 6. Recombination depends on structure and species

If the degree of heterozygosity of two HIV-1 genomes, copackaged within an infectious retrovirus particle, is insufficient to generate sequence differences of a magnitude that inhibits recombination and fosters speciation (Section 7), then, by virtue of similarities in sequence and structure, the two genomes can generate recombinants with properties more advantageous than those of the originating parental genomes (Simon-Loriere et al., 2011). But HIV-1 itself, by virtue of its high mutation rate, explores sequence space in years, to an extent that would take its host millions of years. Thus, as sequence differences increase, a differentiation of HIV-1 into "quasispecies" (up to approx. 15% intrasubtype variation), "subtypes" (approx. 15% - 30% intersubtype variation) and "groups" (>30% intergroup variation), is recognized. How efficiently can recombination have checked this variation? Can differences in recombination between members of these various categories, although mechanistically copy-choice (template switching without strand breakage), guide our understanding of speciation in more complex organisms where recombination is likely to involve DNA strand breakage (Forsdyke, 1995, 1996)?

Recombination in HIV-1 begins with the formation of RNA dimers by means of complementary loop-loop "kissing" interactions between "dimer initiation sequences" (DIS) that are part of stem-loop secondary structures in the Ψ (packaging signal) region of the genome. A similar DNA "kissing" may be involved in recombination in higher organisms (Forsdyke, 1996; Danilowicz et al., 2009). A lack of base complementarity between HIV-1 DIS loops decreases recombination, but does not eliminate it. This suggests that other parts of the genome can assist dimer formation. Upon transfer to a new host cell, copy choice recombination occurs between the colocalized genomes in HIV-1 homodimers or heterodimers (Onafuwa-Naga and Telesnitsky, 2009; Nikolaitchik et al., 2011).

Since a species is defined by its recombinational (reproductive) isolation relative to other species (Forsdyke, 2001), it is important to note that, albeit rare, recombination can occur between members of different HIV-1 groups (Takehisa et al., 1999). This defines them as belonging to the same species. On the other hand, although HIV-1 and HIV-2 can coinfect an individual, no recombinants have been detected, so they are different species. HIV-1 intrasubtype recombination is more frequent than intersubtype (intragroup) recombination, which is more frequent than intergroup recombination (Chin et al., 2007). Thus, after introduction of synonymous substitutions, it was found that 5%, 9%, and 18% sequence differences corresponded, respectively, to 35%, 74% and 95% decreases in recombination frequency (Onafuwa-Naga and Telesnitsky, 2009). Others, with a different assay, found 9% and 18% sequence differences corresponded to 67% and 90% decreases in recombination (Nikolaitchik et al., 2011). These values are consistent with previous FORS-D studies (Section 7), which postulated that HIV-1 secondary structure is conserved because, in addition to regulatory roles, it plays a critical role in recombination (Forsdyke, 1995). Consistent with this, Simon-Loriere et al. (2011) note:

> "Strong disparities [in recombination] were observed for comparable degrees of sequence identity in conserved regions, indicating that … parameters other than the level of sequence identity modulate … recombination. … [R]egions of the genomic RNA with a high proportion of residues involved in the formation of secondary structure contained significantly more [recombinational] breakpoints. The extent of RNA structure along the HIV genome seems to provide us with a relatively accurate picture of the pattern of recombinant genomes generated by the mechanism of recombination."

**7. Base order conserves structure**

As the differences between two subtypes increase, it becomes evident that conservation (low substitution rate) corresponds to regions where bases are ordered to support the formation of higher ordered local structure (i.e. the base order-dependent component of stem-loop potential is maximized). This is shown in Figure 1 where subtype HIVSF2 differs by 455 dispersed substitutions (4.68% difference) from the reference sequence (HIVHXB2). Here structural stability in 200 base moving windows (expressed in negative kilocalories/mol), is plotted with the corresponding substitution frequency for each window. Where the base order-dependent folding stability is high, substitutions are low. Where base order-dependent folding stability is low, substitutions are high. In other words, base conservation associates with stable RNA structure.

This relationship is better seen when the two values within each window are plotted together. The reciprocal relationship does not hold for the base composition-dependent component (Fig. 2b), but does hold for the base order-dependent component (Fig. 2c). Although the points are widely scattered, approximately 10% of the decline in the base order-dependent component of the folding energy is accounted for by increasing substitutions ($r^2 = 0.098$). This result, together with the results of studies with other HIV-1 pairs whose sequences differed more or less than in this case, are summarized in Figure 3 (adapted from Table 1 of Forsdyke, 1995). When differences between HIV-1 genomes are low (e.g. 0.77%; 75 substitutions) a correlation between stable structure and conservation is difficult to demonstrate (slope value 0.142, which is not significantly different from zero). However, when differences are intermediate (4.68%; 455 substitutions; see Figs. 1 and 2) the correlation is significant (slope value -0.34; $P < 0.001$). Thus, in a sequence window where bases are conserved (low substitutions), the bases are likely to contribute positively, by virtue of their ordering, to RNA structure. Conversely, in a sequence window where base order is variable (high substitutions) the bases do not contribute to RNA structure (or can sometimes contribute negatively; i.e. positive FORS-D values).

With higher differences between genomes (8%, 12%), slope values decline and are of marginal significance. This implies that the additional substitutions are now entering windows corresponding to conserved regions, where the substitutions can change both base order and composition, so modify RNA structures that, when substitutions were less, would have supported intra-species recombination. Thus, there is a difference threshold above which recombination, with its associated rejuvenating effects (Section 1), begins to fail.

**8. Base composition and phylogenetic analysis**

Above the threshold the potential to follow a new evolutionary path – speciation potential – increases. In this circumstance base composition plays a more crucial role, a very small fluctuation in GC% being able to substantially change structure so that ‘kissing’ interactions fail and speciation can initiate (Forsdyke 1998, 2007). Once species are established, in phenotypically more complex organisms GC% values may then turn to other roles, but in viruses that have not undergone extensive phenotypic adaptation this seems not to occur, and ‘echoes’ of the originating GC% differences may remain (Section 3; Forsdyke, 2001). Such differences have been found useful in the phylogenetic analyses of retroviruses (Bronson and Anderson, 1994).

Likewise, for influenza viruses Sampath et al. (2007) reported that different "evolving virus species" can be differentiated on the basis of GC% differences: "Base composition derived clusters inferred from this [phylogenetic] analysis showed 100% concordance to previously established clades.” Analyses of influenza virus RNA structures show that, as in the case of HIV-1, conserved RNA structure is in potential conflict with other functions. For example, Moss et al. (2011) found for influenza virus that "RNA structural constraints lead to suppression of variation in the third (wobble) position of amino acid codons."

While GC% differences usually do not suffice for phylogenetic analysis in more complex organisms, palindrome frequencies can be informative (Lamprea-Burgunder et al., 2011). Higher ordered nucleic acid structure depends on appropriately placed, complete or incomplete, palindrome-like inverted repeats of distinctive base order and composition. The frequency of short palindromes has low intra-species variance, but high inter-species variance. Mutations sufficient to generate the large variances are sometimes cryptic, in that there are no obvious phenotype differences (Forsdyke, 2013). Yet, such palindrome frequencies can be used to distinguish species.

## 9. Gene definition and recombination

It is suggested that the potential to adopt a higher order structure relates to recombination (Forsdyke, 1995; Simon-Loriere et al., 2010, 2011). Conserved structure may influence where initial strand-switching occurs when there is copy-choice recombination, and where initial crossovers occur in conventional recombination (Forsdyke, 1996; Smyth et al., 2012). Recombination is also linked to the seemingly endless debate on how to define a gene (Forsdyke, 2009). There is an apparent discrepancy between the gene as defined by biochemists and the gene defined by G. C. Williams and R. C. Dawkins as a “selfish” element that is able to resist recombinational disruption. Evidence that final recombinational crossovers are preferentially located close to gene boundaries brings the two definitions into close correspondence (Forsdyke, 2011b). Studies of the HIV-1 genome further support this. A study of intergroup recombination by Takehisa et al. (1999) revealed that:

> “Breakpoints appeared mostly near the boundaries of the respective genes. The high frequency of recombination that occurs only near the beginning or end of the respective genes seems to reflect a common adaptive strategy for recombination.”

As noted above (Section 5), SHAPE analysis indicates that conserved structure corresponds to the interdomain regions of various polyproteins (Watts et al., 2009). Further application of SHAPE led Simon-Loriere et al. (2010), to conclude that:

> “Junctions between genes are enriched in structured RNA elements and are also preferred sites for generating functional recombinant forms. These data suggest that RNA structure-mediated recombination allows the virus to exchange intact genes rather than arbitrary subgene fragments, which is likely to increase the overall viability and replication success of the recombinant HIV progeny.”

Smyth et al. (2012) agreed:

> "As junctions between genes are enriched with RNA structure … one could argue that the HIV genome has evolved to exchange intact genes as genetic units, rather than as random fragments of the genome, which should increase the chances of recreating a viable virus."

## 10. Conclusions

Having evolved the strategies both of extreme purine-loading and of extreme mutation to evade host defences, *in extremis* only HIV-1 characters that support recovery from mutation (i.e. diploidy, recombination) need to be conserved. Thus the observation of high conservation of RNA secondary structure in HIV-1 that is particularly dependent on local base order, indicates a critical role for structure in recombination. However, when mutational differences exceed a certain threshold, changes in base composition (GC%) can impair structure, and hence impair recombination. Many observations can be explained on this basis, including that failure of recombination (known as 'reproductive isolation') that can lead to speciation (Forsdyke, 2013). Mechanisms of viral speciation may be of heuristic value for studies of speciation in more complex organisms.

## Acknowledgements

My AIDS studies were supported by the American Foundation for AIDS Research and the Medical Research Council of Canada. Queen's University hosts my evolution education webpages where some of the cited references may be found (http://post.queensu.ca/~forsdyke/evolutio.htm).

## Figures

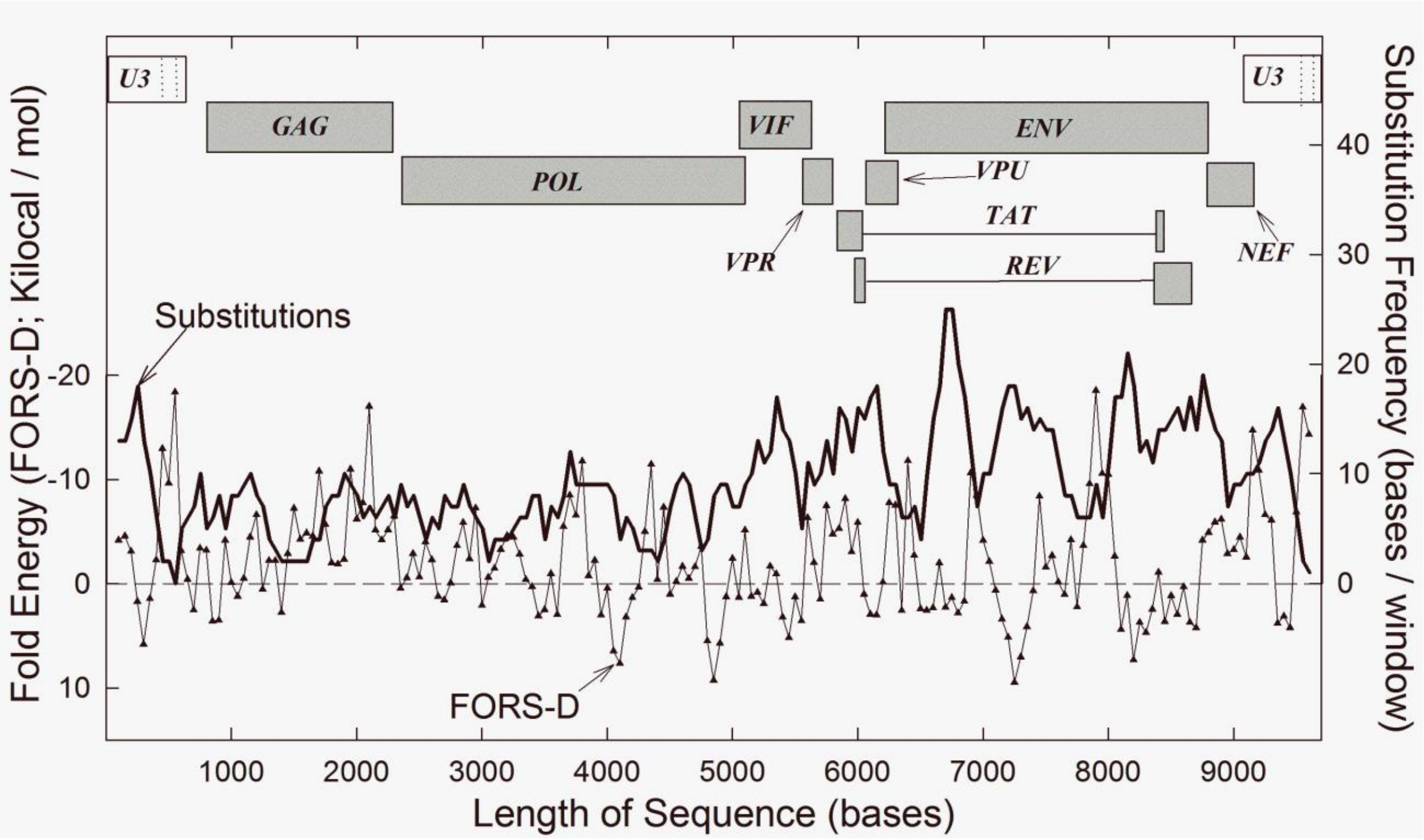

**Fig. 1**. Relationship between HIV variation (base substitutions; thick line) and the base order-dependent component of RNA structure stability (FORS-D; triangles). Genes are shown as gray boxes with their abbreviated names. Base substitutions (total 455) for virus subtype HIVSF2 are relative to the reference subtype HIVHXB2. All values are for 200 base overlapping windows moved in steps of 50 bases. Reproduced from Forsdyke (1995), except that FORS-D values are calculated by *subtracting* the base composition-dependent component of the folding energy (FORS-M) from the total folding energy (FONS). Thus high base order-dependent stem-loop potential (FORS-D) corresponds to high negative values (kcal/mol).

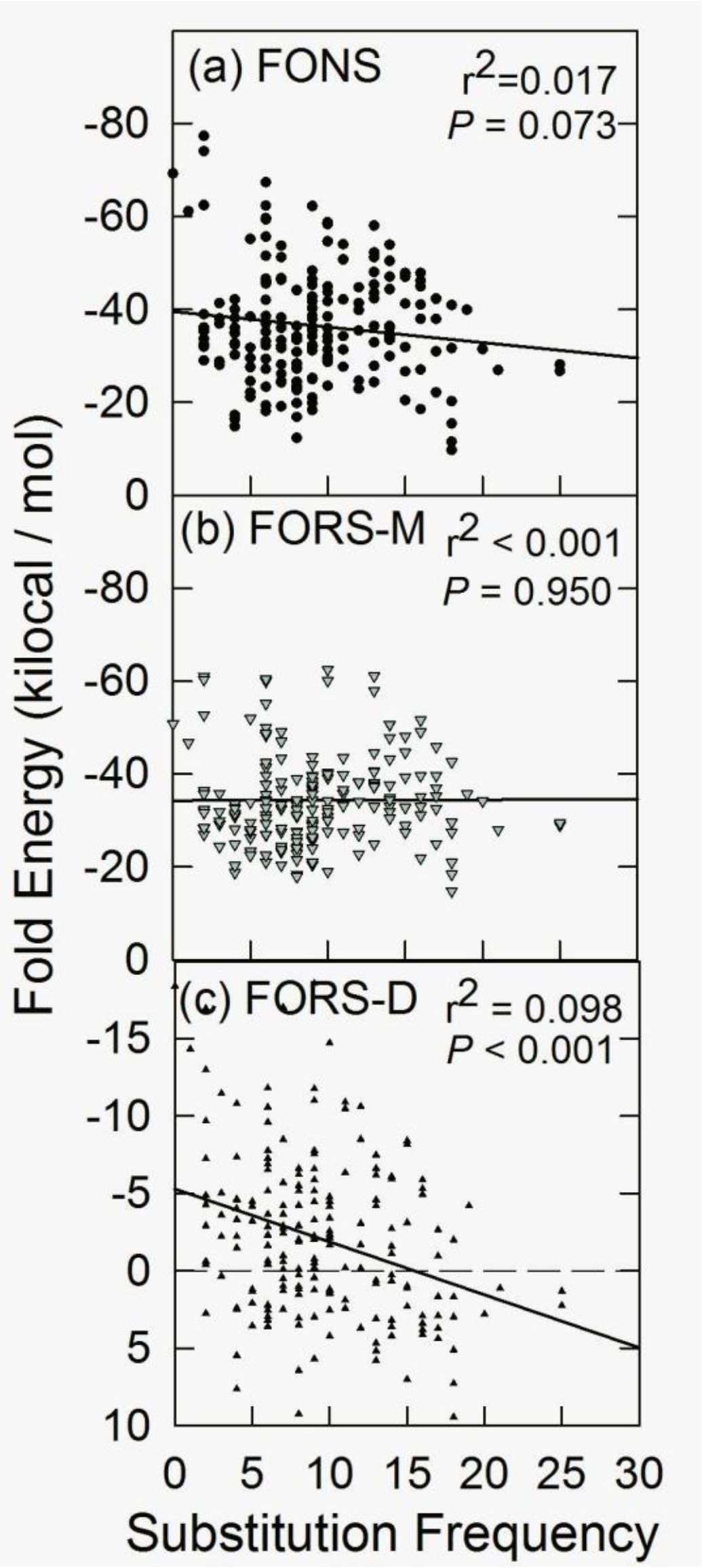

**Fig. 2**. For HIV there is a reciprocal relationship between the base order-dependent component of the stem-loop potential (FORS-D) and substitution frequency in a sequence window (c). But there is no detectable relationship between the base composition-dependent component of the stem-loop potential (FORS-M) and substitution frequency (b). Details are as in Figure 1. Shown in (a) are values for the total folding energy (FONS), which are the sum of FORS-D and FORS-M values in each sequence window. Linear regression parameters indicate that approx. 10% of the variation in substitutions can be explained by base order-dependent stem-loop potential (shown by the $r^2$ value in (c)), and the downward slope is significantly different from zero slope (shown by the *P* value in (c)).

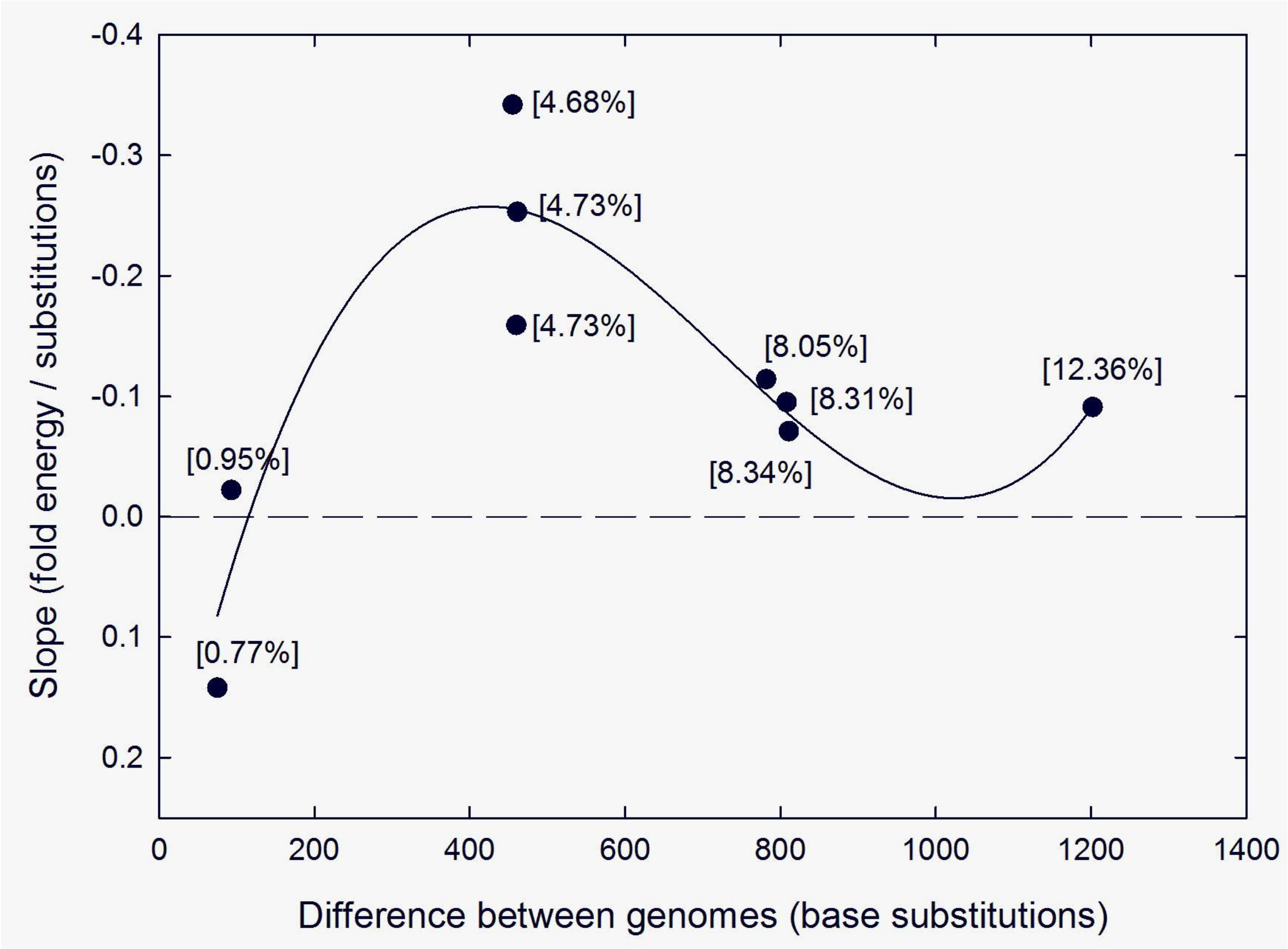

**Fig. 3.** Change in the relationship of fold energy to substitution frequency (e.g. slope of plot shown in Figure 2c) for various HIV isolates that differ from the reference isolate by increasing numbers of base substitutions (expressed as percentages in parentheses). Data, fitted to a cubic polynomial, are adapted from Table 1 of Forsdyke 1995. For further details see text.